\documentclass[a4paper,useAMS,usenatbib]{mn2e}
\usepackage[final]{graphics}
\usepackage{epsfig}
\usepackage{amssymb}
\usepackage{dblfloatfix}

\title[The MarkI helioseismic experiment.I]
{The MarkI helioseismic experiment.I. Measurements of the solar gravitational redshift (1976-2013)}
\author[T. Roca Cort\'es and P. L. Pall\'e]{T. Roca Cort\'es$^{1,2}$  and P. L. Pall\'e$^{1,2}$\thanks{E-mail:
trc@iac.es; pere.l.palle@iac.es}\\
$^{1}$Instituto de Astrof\'isica de Canarias (IAC). E-38200, La Laguna. Tenerife. Spain\\
$^{2}$Departamento de Astrof\'isica. Universidad de La Laguna (ULL). E-38206, La Laguna. Tenerife. Spain\\}

\begin{document}

\date{Accepted 2014 June 20.  Received 2014 June 19; in original form 2014 February 21}


\maketitle

\label{firstpage}

\begin{abstract}
The resonant scattering solar spectrophotometer `Mark-I', designed and build at the University of Birmingham (UK)
and  located at the Observatorio del Teide (Spain), has been continuously in operation 
for the past 38 years. During this period of time, it has provided high precision 
measurements of the radial velocity of 
the Sun as a star, which has enabled the study of the small velocity fluctuations produced by 
the solar oscillations and the characterization of their spectrum. So far, it has been one of 
the pioneer experiments in the field of helioseismology and contributed to the development of that area.
Moreover, because of its high 
sensitivity and long term instrumental stability, it also provides an  accurate determination (to within a few 
parts in $10^3$) of the absolute daily velocity offset, which contains the so-called `solar gravitational 
red-shift'. 
In the present paper, results of 
the analysis of the measurements of this parameter over the whole period 1976--2013 are presented. 
The result of this series of measurements is 600.4 $\pm$ 0.8 m s$^{-1}$ with an amplitude 
variation of $\pm$5 m s$^{-1}$, which is in anticorrelation with the phase of the solar activity cycle. 
The 5\% difference found respect to  the value predicted by the equivalence principle is probably due to the asymmetry of the solar spectral line used.
\end{abstract}


\section{Introduction}

The redshift of spectral absorption lines from the solar photosphere has been 
known empirically since 1896 \citep{Jew1896} and was then interpreted in terms 
of pressure shifts in the plasma of the solar photosphere. The formulation 
of the principle of equivalence and later of 
the General Theory of Relativity \citep{Ein1911, Ein1916} indicated that this phenomenon 
was due to the difference in gravitational potentials between the solar photosphere 
and the Earth's surface. This encouraged extensive observational work that has 
established with certainty that the solar redshifting of spectral lines is quite a complex 
phenomenon with the posibility of velocity fields, collisional efects, magnetic 
effects and other factors adding to the gravitational redshift (hereafter GRS) \citep{Adam1948, 
LoPresto1994,Caccianietal2006}.

In contrast with astrophysical applications of the GRS, an experimental test became 
possible only after the discovery of the M\"osbauer effect \citep{Pound&R1960} and 
is undoubtedly capable of substantial improvemente in the future. Moreover, the use of
atomic clocks \citep{Vessot1979, Vessotetal1980} has improved the accuracy of the measurements.
There is no doubt that better clocks, as well as laser-based frequency standards,  
 will make 
further improvements possible of the accuracy of measurements of the GRS 
on the Earth's surface or in near-Earth space in the near future. Moreover, the use of 
laser-frequency combs is becoming a possibility for astronomical observations too \citep{Steinetal2008}.

There has also been progress in our understanding of the conditions any 
theory has to satisfy in order to account for the GRS and also the 
implications of the existence of the GRS. These considerations 
show that a wide class of theories predict an identical GRS, and it is 
now appreciatted that it is a rather weak test of relativistic theories of gravitation.
However, \cite {Sciama1964} points out that a measurement of the solar GRS, 
being over a distance very much greater than the scale height of the 
gravitational field is in principle preferable to local laboratory 
experiments and can test the validity of the minimum coupling principle.

In the final decades of the 20th century several measurements of the solar GRS were made 
using various techniques, such as improved solar spectrographs \citep{Brault1962,LoPresto1991} and
 atomic beams \citep{Isaak1961,Snider1970,Snider1972,Brookes1974}.
 The reported agreement with theoretical predictions for the strong resonance 
absorption lines of potassium and sodium, which are formed high in the photosphere,
 are within observational error of some 10\%, including systematic effects. 
More recently  \citep{LoPresto1994,Caccianietal2006}, 
there has been a better comprehension of the solar effects that add to 
the GRS, thereby reducing systematic errors and improving the  results obtained.

Moreover, \cite {Brookes1974}, \cite{Isaak1976} and \cite{Fossat1973} introduced the 
use of alkali-vapour cells based resonant scattering spectrophotometers in the 
measurement of radial velocities in the Sun's 
photosphere when observed as a star, resulting in an improvement of more 
than an order of magnitude over existing techniques \citep{Jimenetal1986}. 
Works based on this technique, and some others, led to
 the discovery of solar oscillations and the birth of helioseismology. 
One such key instrument \citep{BIR1978}, named MarkI, designed and built at the University of Birmingham (UK), 
has been in operation since 1975 at the 
Observatorio de El Teide (Canary Islands, Spain).

The dedication of this apparatus for such a longtime
to solar observations has provided an enormous amount
of valuable data and moreover, it served as a reference for
other similar instrumentation build and operated later on,
both in ground and in space. Now, with some historical perspective
too, in a series of two papers we summarize the
long standing observations and results achieved, we analyse
the whole data obtained till now and report new findings
and discoveries resulting from the precise measurements of
the Doppler shifts of the 7699 \AA resonance line of neutral
potassium atom in the light integrated over the entire Sun
(the sun viewed as any other star).

In this first paper, we briefly
describe the apparatus, the observations achieved and the results of the
analysis of its long-term time stability velocity signal which
provided a long series of GRS measurements spaning from
the years 1976 to 2013. In the second paper, a full in-depth description
of the calibration of the data will be given and results of the analysis of the
spectrum of the oscillations of the sun along three full solar 
activity cycles will be presented.

\section{Principle of the method and apparatus}

\begin{figure}
\includegraphics[width=\columnwidth]{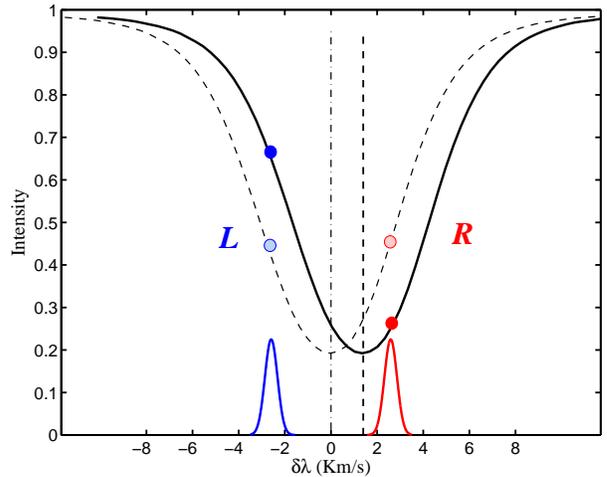} 

\caption{\label{Figlines} Principle of operation of the vapour cell resonant 
scattering spectrophotometry (see text). The relative displacement of the solar 
line with respect to the laboratory ones can be
calculated by measuring at both wings (R and L) and making an appropriate ratio between 
both. In our case, observing integral sunlight with KI7699\AA \  line, the 
dynamical range of the velocity shift spans from -300 m s$^{-1}$ (end September) to 1500 m s$^{-1}$ (beginning April).}
\end{figure}

The method used by resonant scattering spectrophotometry has already been 
described in detail \citep{BIR1978} and may be summarized here using Fig.\ \ref{Figlines} as a 
reference. Circularly 
polarized light with an absorption line, represented by the solid 
black curve, is incident on a suitable atomic vapour cell having 
a resonance line overlapping with the solar absorption line, placed in a 
longitudinal magnetic field of such strength as to locate the 
Zeeman components (blue and red curves) near the steepest parts 
of the solar absorption line. The intensity of resonantly scattered light 
due to left-handed circularly polarized incident light is given by $L$, 
whereas that due to right-handed circularly polarized light 
is given by $R$.  It is clear from   Fig.\ \ref{Figlines} that, when there is no 
relative displacement between the incident spectral line (dashed line) and the 
laboratory lines, $L$ equals $R$. If, however, there is a relative 
shift between both lines, as illustrated in the figure (solid line), the two 
intensities are no longer equal, and the ratio defined as

\begin{equation}
\label{rati}
 r=\frac{L-R}{L+R} \propto V_{OBS}
\end{equation}

gives a measure of the line shift corrected for intensity 
fluctuations to first order. Since the narrow laboratory Zeeman 
components scan the steepest parts of the broader solar line as the Earth spins, this 
method is extremely sensitive to very small shifts. In this 
experiment natural potassium vapour was used to study the relative position
of the KI7699 \AA \  absorption line with respect to the observatory. The solar line has a full width at half depth of 
6.23 km s$^{-1}$, while the region probed by the laboratory lines is of order of
1.8 km s$^{-1}$, thus the ratio $r$ will be almost linearly related to the 
velocity shift between the Sun and the laboratory to a first order 
approximation; in fact, a constant calibration of 3000 converts the ratio 
into velocity in m s$^{-1}$ \citep{Bobetal1986}. A more detailed account of modifications due to the 
hyperfine structure and isotopic composition can be found in 
the more detailed and comprehensive paper \citep{BIR1978}.

The MarkI apparatus, was developed and built at Birmingham 
University (UK) during 1973 and 1974, and was gradually improved 
over the following five years. A detailed account of its state 
during 1976 and 1977 in its first location in the \textit{Casa Solar} 
at the Observatorio de El Teide (Tenerife, Spain) is to be found in \cite{BIR1978}
and \cite {Teo1979}.  Briefly, 
a small, equatorially mounted servo-controlled heliostat (with flat mirrors)  
directs disc-integrated sunlight, suitably filtered by a 
 thermostatically controlled interference filter 
centred at 7699 \AA\  with a bandwidth of 15\ \AA, through a polarizer  
and an electro-optical light modulator, which alternately produces the required 
left- or right-handed circular polarization by the application of 
appropriate electric potential. The light then enters a vapour 
cell situated in a longitudinal field provided by a permanent magnet (0.18 T); the resonantly 
scattered light is then detected by a cooled photomultiplier tube (PMT), and 
the processed output pulses are recorded  
for subsequent computer analysis. In order to test for various 
optical and electronic asymmetries and to assess the overall stability of the 
system, a light source on its own or a white light (quartz-iodine) 
with a vapour cell in a transverse magnetic field to provide an 
artificial absorption line simulating the solar potassium line was also 
used.

\section{Observational procedure and strategy}
\label{obsproc}

From 1975 onwards, the MarkI instrument has been used exclusively 
to study the oscillations of the Sun as a star, and the measurement
 of the GRS is a byproduct that will be studied in this work.
The one and only exception was in the 1979 summer campaign, in which 
the apparatus was used to look at different parts of the solar disc.
 Integrated sunlight helps to reduce the random shifts of the spectral
 line due to small scale motions in the solar photosphere.
 Moreover, it has the advantage that the rotation of the Sun is averaged
 to zero and possible instrumental errors due to imprecise guiding or 
atmospheric turbulence become minimized. Small signals due 
to angular non-uniformities on the photosphere, such as sunspots, or in 
sky transmission, or instrumental errors will be discussed later on.

\begin{figure}
\center
\includegraphics[width=\columnwidth]{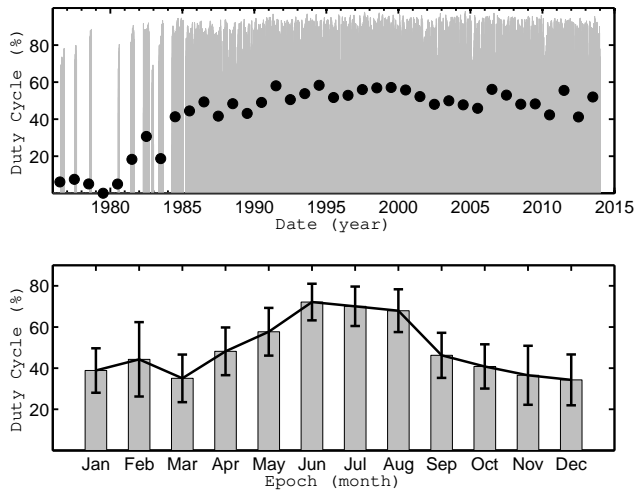} 
\caption{\label{Figyears} Top: duty cycle of daily MarkI solar observations throughout the 
years in percentage of daily useful hours (grey bars); the black dots 
are the annual values achieved in that year. Bottom: monthly averages
(over the observed years) of the duty cycle with its standard deviations.
}
\end{figure}

MarkI was taken to the Observatorio de El Teide in 1975 and began operation
 by the end of that 
summer for only a few days. Several long summer observing campaigns 
were then carried out there until 1984. During those years MarkI's performance 
was being monitored and used to develop other apparatus, based on 
the same principle, that helped to establish helioseismology as an 
astrophysical discipline in its own right. Moreover, from summer 1985 onwards, 
the observations were taken on a continuous daily basis in a fairly undisturbed 
and regular observational procedure. Its precise location at the 
observatory has been changed twice, the most recent being in 1986 when
it was moved to its current location, the solar lab named \textit{Pir\'amide van der Raay}.
From 1993 onwards, it is also a node of the BiSON helioseismology network \citep{Chaplin1996}.

The sunlight enters the spectrophotometer through an 
open-air double flat mirror heliostat system. The two-axis servo-controlled 
primary mirror (which also moves along the north--south line throughout the year) 
feeds a fixed secondary mirror, which provides a horizontal sunlight 
beam towards the spectrophotometer room. This system, although symmetric about 
noon, has to be altered by changing the secondary mount at the equinoxes, 
otherwise it will cast shadow on to the primary around noon. Before 1979, 
the heliostat system was different and had an extra fixed flat mirror to accommodate the sunlight 
beam properly; also, the mirrors were of lower quality.

Minor changes in the core parts of the apparatus have been introduced, 
and it can be said that, up till 2009, MarkI has been working without 
interruption, weather permitting, other than a few  failures that have 
been solved quite rapidly. The main changes have been in the electronic readout 
system that evolved from fast analogue electronics and 
seven-track magnetic tapes to fast digital electronics and a microcomputer 
system with cassette tapes in 1984; obviously, new computer hardware and software has been
incorporated to the apparatus as it has become available and possible. Its performance has been outstandingly 
good; just recently,  in October 
2009, the vapour cell and PMT had to be changed because of darkening 
of the windows in the former and loss in efficiency in the latter, which
lowered the count rate and signal-to-background ratio.

The raw data consist of a one-second measurement on the left-hand side 
of the line followed by a second on the right hand. 
The count in each channel usually spans from 0.1 to 1 $\times 10^6$
counts per second depending on sky transparency, thin cirrus and mirror cleanliness. 
Furthermore, the ratio is calculated by finding its 
average and standard deviation  on the basis of blocks of 42 s (40 s from 1984 on),
 which, when calibrated to velocity, results in an error of $\approx$ 1 m s$^{-1}$.

The census of the observations during these years can be seen in 
Figure~\ref{Figyears}. Longer runs have been achieved in summer months when a higher number
of daytime hours and cloudless days is more frequent; June, July and August are consistently 
the best months of the year. However, in winter 
the duty cycle is consistently lower even though the transparency 
of the sky is better; December and March are the months with the lowest duty cycle (around 40\%). Note
that, up to 1985, observations were taken only in summer-month campaigns. The duration of
failures due to instrumental problems would be equivalent to a couple of useful days per year on average.

To summarize the amount of data in hand in a few numbers, it may be said 
that in the period from July 1976 to December 2013, with a total span of 
13672 days, we have observed for 8849 days. However, in the period from June 1984 
to December 2013, where daily observations are available, out of a total span of 
10958 days we obtained 8390 days of observations for a duty cycle of 49\% of all possible daily hours.

Currently, the whole MarkI (level 1) database is accessible at the SVO site 
(Spanish Virtual Observatory  http://svo2.cab.inta-csic.es/vocats/marki).

\section{Data analysis}
\label{sec:analisis}

The result obtained on a typical day of observation is shown in Fig.~\ref{Figdayobs} (top graph).
If all the displacements of spectral lines of the Sun with respect to the 
laboratory are expressed in terms of relative velocity of the Sun with 
respect to the laboratory, at any given time we have:

\begin{figure}
\center
\includegraphics[width=\columnwidth]{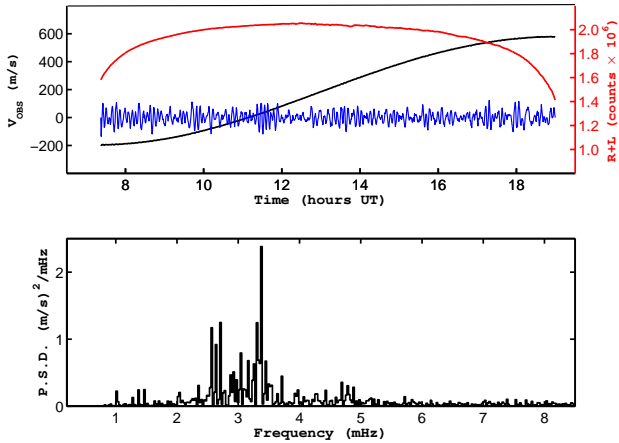} 
\caption{\label{Figdayobs} Results of a typical day's observing with MkI (2011 August 30).
The top graph shows the following curves: observed velocity (in black), a 
hundred times the residuals after a sine 
wave fit (in blue) and the light transmitted (in red) through the apparatus. 
The bottom graph shows the power spectral density of the residuals
 with the 5 min solar oscillations being clearly seen above noise level.
}
\end{figure}

\begin{figure}
\center
\includegraphics[scale=0.45]{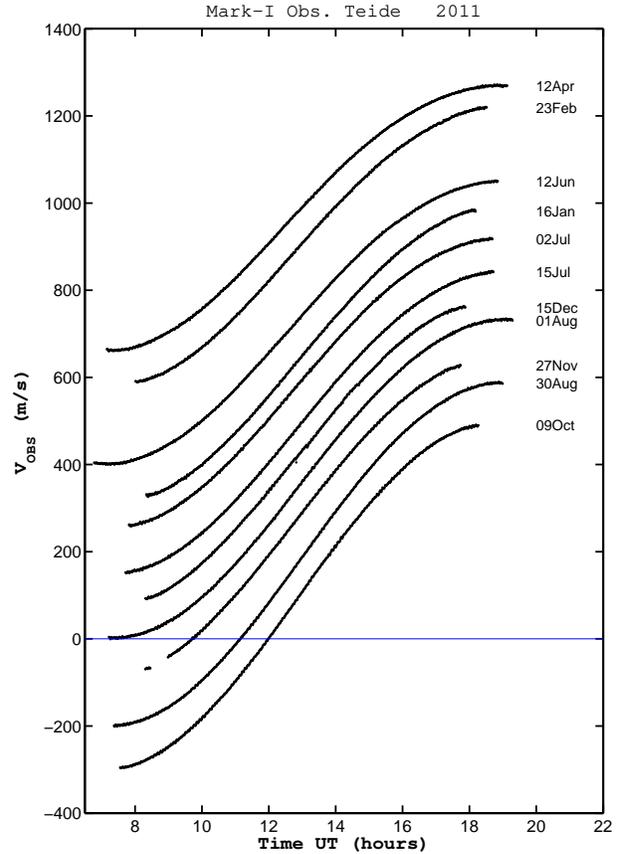} ~ 
\includegraphics[scale=0.65]{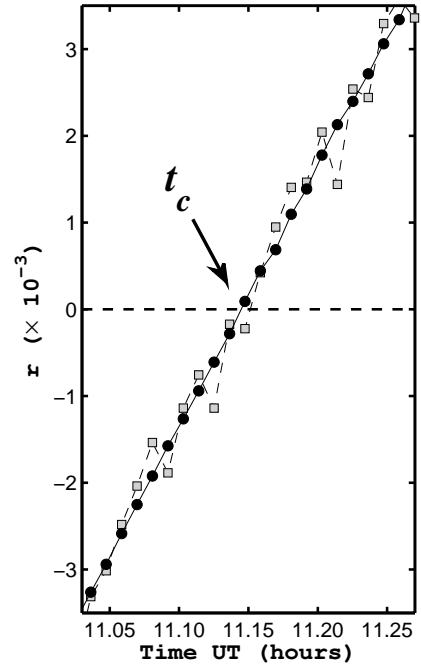} 
\caption{\label{Figanual} Top graph: a few  days' observations spread 
over the year, illustrating that only around the interval spanning
from the beginning of August to the beginning of December does the observed daily 
velocity curve cross the zero value. Bottom graph: zoom to show the obtaining of the 
crossing time, $t_c$; that is, the precise time of the day when the velocity of 
the sun relative to the observatory is zero (on 2011 August 30).
}
\end{figure}

\begin{equation}
\label{eqVobs}
V_{OBS} =  V_{ORB} + V_{SPIN} + V_{GRS} + V_{0}+V_{OSC} \ \,
\end{equation} 
where,
\begin{equation}
 \label{eqvorb}
V_{ORB} = (\bf{V_{\odot}}-\bf{V_{\oplus}}) \cdot \bf{u_r}  
\end{equation} 
is the resolved component of the relative velocity between the Sun and the Earth 
along the unit radius vector joining their centres, with due allowance 
for the planetary and lunar perturbations,
\begin{equation}
\label{eqvspin}
V_{SPIN} = {\bf V_{rot \oplus} \cdot {u_r}} = \Omega_{\oplus} R_{\oplus} \cos \lambda_{obs} 
\cos \delta_{\odot} \sin \left [\frac{\pi}{12} (t-t_0) \right]
\end{equation}
is the component due to the observatory's daily rotation, where 
$\Omega_{\oplus}$ is the angular velocity of the  Earth, $R_{\oplus}$ 
is the observer's distance from the centre of the Earth, $\lambda_{obs}$ 
is the latitude of the observatory and $\delta_{\odot}$ is the declination 
of the Sun and $t_0$ is the local noon, 
\begin{equation}
\label{eqvgrs}
V_{GRS} = \frac{GM_{\odot}}{c} \left(\frac{1}{R_{\odot}}-\frac{1}{2 \cdot 1AU} \right)-
\frac{GM_{\oplus}}{c R_{\oplus}} = 633.7 \ \mathrm{ m\  s^{-1}}
\end{equation}
is the GRS velocity equivalent (including the relativistic Doppler effect), where $G$ is the gravitational constant, 
$c$ the speed of light and $M_{\odot}, R_{\odot}, M_{\oplus}, R_{\oplus}$ 
are the masses and radii of the Sun and Earth respectively, and
\begin{equation}
 \label{eqvsun}
V_0 + V_{OSC}(t)
\end{equation}
are unknown solar velocities; the first term representing the aperiodic solar velocity 
fields that can be considered stable over a day while the second stands for 
the measured global oscillations with periods lower than a day \citep{Clavetal1979}.

Figure~\ref{Figdayobs} shows the output of a typical day's observing, where
the ratio $r$ (in black, already calibrated in velocity), 
its residuals after a sine wave fit (multiplied by 100, in blue)
and the transmitted light (in red) through the apparatus are shown; the bottom graph of 
Fig.~\ref{Figdayobs} shows the power spectrum of the residuals of a fitted sine wave to the ratio/velocity curve,
 where the resulting solar five minute oscillations are clearly visible.

The term $V_{GRS}$ is expected to be constant (over an Earth orbit it will 
only change by as much as 0.1 cm s$^{-1}$), 
$V_{ORB}$ varies along the 
year, from 502 m s$^{-1}$ around April to -500 m s$^{-1}$ around October, 
with a maximum of $\approx$ 12 m s$^{-1}$ between successive days, whereas the 
$V_{SPIN}$ varies sinusoidally over a day with an amplitude that changes 
from 375 and 408 m s$^{-1}$ at Observatorio de El Teide, depending on the Sun's 
declination. These terms can be better seen in Fig.~\ref{Figanual} (top). Note 
that these Earth movements provide an accurate (and daily) calibration 
of the Doppler measurements in terms of velocity units to the nearest cm s$^{-1}$. 
On the other hand, instrumental, systematic and random noise can provide 
spurious velocity terms that will be dealed with further in this paper. 

\section{Two methods to measure the GRS}
\label{sec:methods}
 
\subsection{Null measurement method}
\label{sec:nullGRS}
The seasonal changes of the $V_{obs}$ daily observations (due to $V_{ORB}$) can 
also be clearly seen (Fig.\ref{Figanual}) 
and it can be noticed how the $r$ ratio crosses  
zero value once at some time in the mornings of summer--autumn seasonal days. 
In fact, at Observatorio de El Teide it happens roughly from  
August 1 to  December 1. At that precise time $t_c$, the solar absorption line 
and the laboratory line have a relative velocity zero shift. It is clear that 
such a null observation provides a very sensitive means of determining 
a daily value for $V_{GRS}+V_0$ (see equation \ref {eqVobs}) in terms of the very accurately 
known value of $V_{ORB}+V_{SPIN}$ at $t_c$(see Fig.\ \ref{Figanual}, bottom) from 
 the astronomical ephemerides. 
The term $V_{OSC}(t)$ with the information of the solar oscillations 
is very small in amplitude ($\approx$ 1 m s$^{-1}$), having low period signals
 (around 5 min); obviously, this signal can be filtered out quite well if need be.
Moreover, this null measurement is independent of various background instrumental 
noise problems and curvature effects of the solar absorption line near the 
operating region, being independent of the calibration of the 
apparatus in terms of velocity, at least to first order.
 
\begin{figure}
\center
\includegraphics[width=\columnwidth]{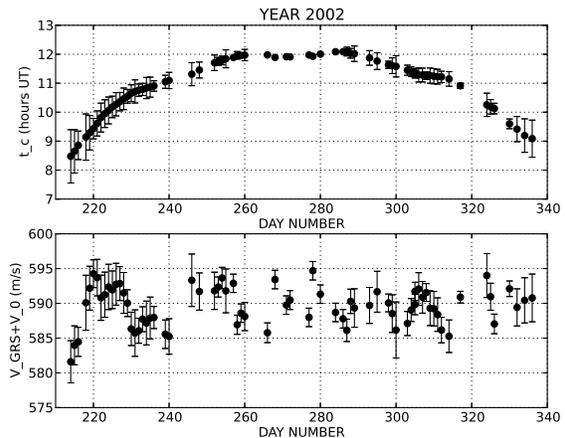}
\caption{\label{Figvgrsanual} The GRS velocity obtained during
any of the observed years. Top, the crossing 
time obtained from the observations (error bars are 10 $\sigma$)  
and, bottom, the corresponding calculated
 velocity with 1$\sigma$ error bars.  
}
\end{figure}
 
\begin{figure}
\center
\includegraphics[width=\columnwidth]{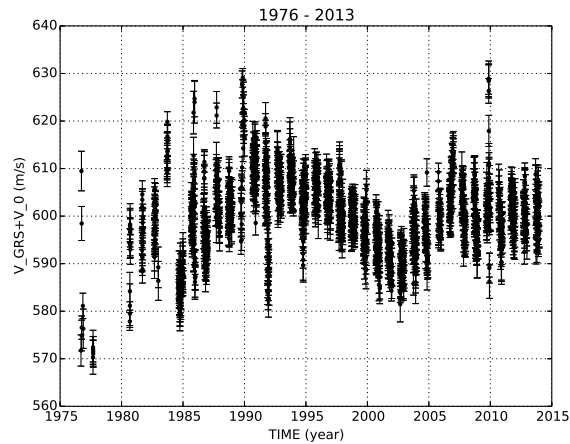} 
\caption{\label{Figrawvgrsanual} Raw values of the GRS velocity obtained during 
all years on the days where observations are available and the crossing time can be measured. 
}
\end{figure}

Therefore, a way of obtaining the crossing time is to fit a straight line to the data 
on a given short time interval around the zero velocity crossing. The interval should be short enough 
so that a straight line is a good approximation and long enough so that the 
oscillation signal can be appropriately averaged out. The interval found to be 
appropriate is 40 min centred on the crossing time. The actual way of measuring 
the crossing time is shown in Fig. \ref{Figanual} (bottom); following the 
use of a moving mean filter, a straight line is fitted and the crossing 
time and its uncertainty calculated. The fit has been made with and without 
filtering; moreover, several appropriate filters were also checked 
without significant changes in the result.

For any one year of observation typically under 
90 daily independent measurements are obtained that can be seen plotted in Fig.\ \ref{Figvgrsanual}, where 
the crossing time over that season is shown (top graph), while in the bottom graph the calculated 
$V_0^{t_c}=V_{GRS} + V_0$ daily values are also plotted. It can be seen  
that the average value in that year, 589.7 $\pm$ 0.3 m s$^{-1}$, is not the one expected from 
the first term alone; therefore,  $V_0$ is not null. Moreover, there are day-to-day  
variations well outside the errors on each measurement. Such daily peak-to-peak variations, 
around 10 m s$^{-1}$ and periods around half the solar rotation 
period (see Fig.\ \ref{Figvgrsanual}), clearly suggest the activity features in the photosphere 
moving across the visible solar hemisphere. Therefore, we need  to understand 
this result, which is rather similar in all years with observed data. 

Moreover, these calculations are done over all the days of observation in which it is possible
to obtain the crossing time
and the results are plotted in Figure\ \ref{Figrawvgrsanual}. In this figure
 the measurements show a mean value of   601.14 $\pm$ 0.15 m s$^{-1}$, 
 which is some 32 m s$^{-1}$  lower than the 
predicted GRS velocity. Therefore, the $V_0$ term will be non-zero owing to  
contributions from several sources, both solar and instrumental. 
 
\subsection{Measurement at noon method}
\label{sec:noonGRS}

A second method to daily measure the GRS would be to perform a measurement at noon time. 
As explained in section \ref{sec:analisis}, notice that when $t = t_0$ the term $V_{SPIN} = 0$ 
and the equation  \ref{eqVobs} can be expressed as:
\begin{eqnarray*}
\label{eqvobser}
V_{GRS} + V_{0} &=& V_{OBS}(t_0) - V_{ORB}(t_0) - V_{OSC}(t_0)\\ 
& \approx & V_{OBS}(t_0) - V_{ORB}(t_0)
\end{eqnarray*}

\begin{figure}
\center
\includegraphics[width=\columnwidth]{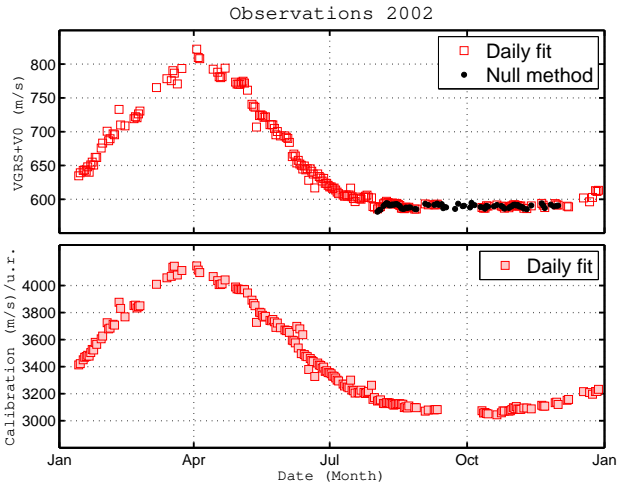}

\caption{\label{Figcali} Results of the offset velocity $(V_{GRS}+V_0$ and the calibration $K$ obtained 
with the noon-measurement method for the year 2002. Notice the different behaviour around April from the 
one around October. Also plotted (black dots) is the result obtained with 
the first method (null measurement method) whenever available (August--November). Notice the 
different behaviour in April, where the most asymmetric parts of the line are sampled, 
from the one around October, where the parts sampled are symmetric.
}
\end{figure}

Therefore, 
a daily value for $V_{GRS}+V_0$ can be obtained in terms of the  
known daily value of $V_{ORB}$ (from the astronomical ephemeries) and the determination 
of $V_{OBS}$ at $t_0$; notice that $V_{OSC}(t_0)$ is lower than 1 ms$^{-1}$. 
However, the determination of $V_{OBS}(t_0)$ involves a process of calibration 
from the instrumental velocity (the observed
ratio, see equation \ref {rati}) to line--of--sight velocity in  ms$^{-1}$. 
On the other hand, this method allows a daily 
measurement along the whole year, provided the method of calibration  is precise and stable enough.
Although a full and precise description of the calibration procedure involving explanations and corrections
of non-linear terms will be done in detail in the second paper of this series, 
we will provide here arguments that prove that 
under the best conditions (those very close to the linear regime) the results 
obtained for the GRS velocity coincide with 
the above explained null method.

A very simple and effective way of daily calibrating the measurements can be
described as follows. From Fig.\ref{Figanual} and equation \ref{eqVobs}, the 
measured ratio at any time can be expressed in the following way:
\begin{equation}
\label{eqrobs}
r (t) = a + b \sin \left [\frac{\pi}{12} (t-t_0) \right]
\end{equation} 
Therefore, assuming a linear relation between ratio and velocity, this equation 
can be fit to the observations of any single day and obtain the daily values of the 
calibration $K$ and the velocity offset $(V_{GRS}+V_0)$ as:
 
\begin{equation}
\label{eqcal}
K = \frac{\Omega_{\oplus} R_{\oplus} \cos \lambda_{obs} \cos \delta_{\odot}}{b}
\end{equation} 
and
\begin{equation}
\label{eqa}
(V_{GRS}+V_0) = K \cdot a - V_{ORB}(t_0)
\end{equation} 

The results obtained for a typical year, say 2002, are shown in Fig.\ref{Figcali}. 
As it can be seen, from beginning of August to beginning of December is when the measurements 
are done under the most favourable conditions (when the most symmetric parts of the line are being 
sampled.see Fig.\ref{Figlines})
and both, calibration and velocity offset, are almost constant; notice that this is precisely the same epoch where
the crossing time method (described above) can be applied.
On the contrary, when measurements are done in the least
favourable conditions, around April when the sampled parts on the wings of the line are 
asymmetric, the calibration changes
by some 35\% and the offset velocity also changes accordingly.
It can also be noticed 
that in this later case the scatter of the points is higher than in the former. Notice
also that the null measurement method, when applicable, coincides extremely 
well with those obtained with this method (the average 
of the differences is  0.6 $\pm$ 0.2 ms$^{-1}$).

In order to better calibrate the apparatus along the whole year it is 
necessary to linearise the measurements (to correct for the non--linearity of the line profile)
in such a way that the results for the calibration become as constant as possible along the whole year. 
Although some methodology for very similar observations
has been extensively described already in the past (see for instance \citep{Bobetal1986, Pereetal1993},
in the next paper of this series it will be fully  discussed.
 
\section{Interpretation and corrections}

Figure\ \ref{Figvgrsanual} shows a clear daily variation with periods around 
half the solar rotation period, the so called 13-day period variation 
due to the passage of sunspots and other inhomogeneities across the visible solar photosphere, 
already pointed out in the past by \cite{Edmonds1983}, \cite{DuSch1983}, \cite{AndMal1983} and \cite{Herretal1984}.
Moreover, the $V_0$ term will, or can, have 
contributions from many other sources, mainly solar but also instrumental. In this
section we will try to understand such contributions.

\begin{figure*}
\center
\includegraphics[angle=270,width=15.0cm]{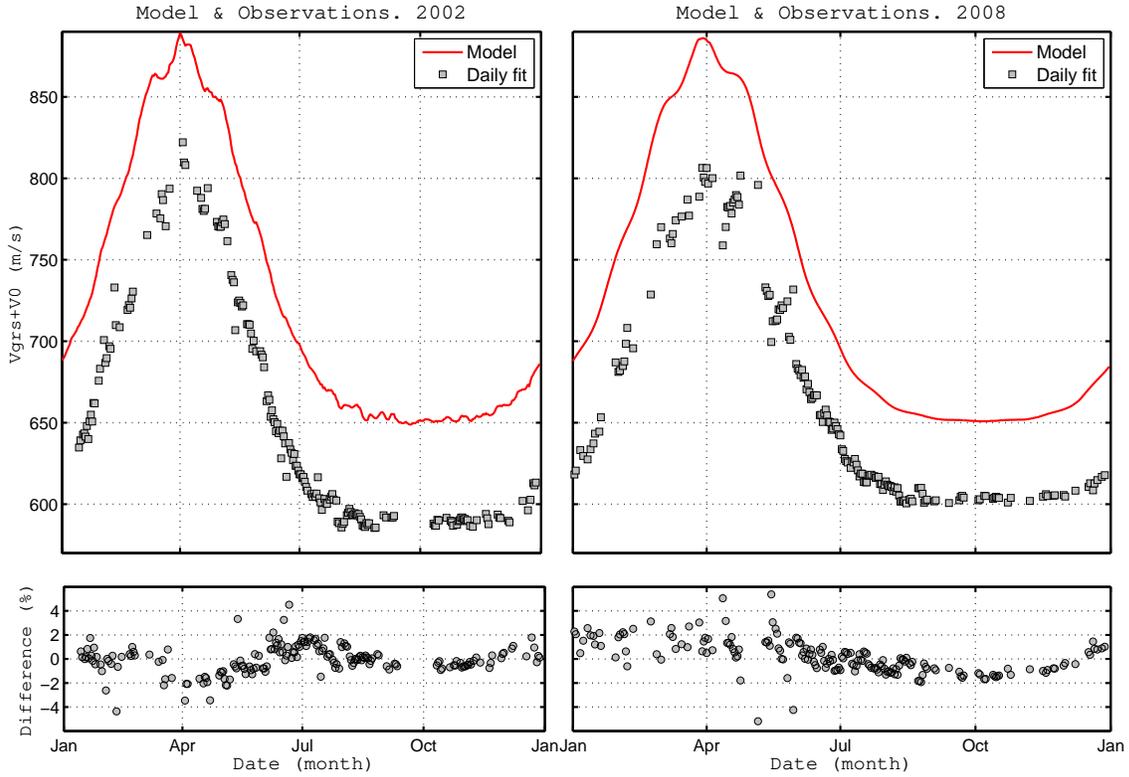} 
\caption{\label{Figanualsimul} Top, results from the numerical model for two years with high (2002) and low  (2008)
magnetic activity and its comparison to observations analysed with the noon-measurement method. 
Bottom, the relative difference is plotted after being substracted its mean value, showing a good 
match and less scatter in the August to November epoch. 
}
\end{figure*}

\subsection{Solar magnetic activity effects}
\label{sec:corsolares}

The signal due to the passage of magnetic activity features over 
the visible solar hemisphere can be modelled by trying to match the observed data and 
can eventually be subtracted from observations. 
A simple model has been developed using 
the impressive sunspot data archive in  NOAA-NGDG (http://ngdc.noaa.gov/),  where daily sunspot 
areas and positions are recorded from well before our observations began
up till now with very few gaps. The numerical 
model is fairly simple and has been used in slightly different ways in the past (see the above-mentioned 
authors). In this paper we have followed the work of \cite {Herretal1984}. 

The numerical model simulates, at any point on the solar surface, all known line-of-sight relative velocities 
of the Sun with respect to Earth and the way the apparatus makes the measurement of the solar lines over the period of a year. 
Data on the potassium solar line, both in the quiet sun and in the sunspots, are taken
from observations in the literature \citep{Rocaetal1983, Bonetal1988, Marmolinoetal1987}. All involved lines
(solar and laboratory)
are taken as gaussians, which is a good enough approximation for the scanning interval
at work in the experiment. Moreover, it also includes the so-called limb-shift effect, 
which is caused by the granulation velocity fields present in the Sun`s 
photosphere whose effect result in  asymmetries in the spectral lines 
(the C-shapes bisectors), and change as we move from centre to limb 
\citep{Dravins1981, Dravins1982, Rocaetal1983, LoPresto1985}. As we observe the Sun 
as a star, the measurements will be affected by the disc-integrated 
result of this effect and for the KI7699\AA \  line 
it has been measured in the past \citep{Angueraetal1987, Andersenetal1985, 
LoPresto1994} and studied theoretically by \cite{Marmolinoetal1987}, turning out to 
be small, with some differences among various authors.
 We have used
an average value of the measured effect from the above mentioned authors, whose
shift of the line is 
approximated by a second order polynomial in $\mu$ ($\mu = \cos \theta$,
where $\theta$ is the heliocentric angle). Finally, it should be stated that the
numerical model designed in this way has no free parameters.

The daily results for years 2002 and 2008 (high and low level magnetic activity years)
are shown in Fig.\ref{Figanualsimul} where the GRS measurements obtained with the 
noon-measurement method (see subsection \ref{sec:noonGRS}) are also shown for comparison.
These simulated results are also calculated for every day at noon and are expressed in velocity using the 
daily calculated calibration from the model. 
As it can be seen, the numerical model follows very well the daily observations on both years, showing
that the model seems to work well, even in both extreme conditions (the most favourable around October
and the least favourable around April). 

The difference between both years (diferent level of solar activity) is in the average value
of the observations (601.5 m s$^{-1}$ and 590.3 m s$^{-1}$ for 2008 and 2002 respectively).
Also, the annually averaged values 
of the simulated data  show a slight difference between both years, being of 0.1 m s$^{-1}$ 
 higher in 2008; other years with the highest activity only showed 1.5 m s$^{-1}$ 
less than the lowest minimum. The offset velocity between model and observations is of some 50 to 60 m s$^{-1}$;
in order to obtain simulated values of GRS close to the ones observed we have to use 
a value for $V_{GRS}$ of 585 m s$^{-1}$ instead the one theoretically
predicted (see section \ref{sec:analisis}) or to modify the limb-shift effect.  
On the other hand, in Fig.\ref{Figanualsimul} the relative difference between model and observations is 
also plotted (mean value is substracted). It is interesting to notice that the flat part of the curve, 
from August to November, is better fitted than the rest of the year where a
higher scatter (more than a factor of two) and some slight curvature is left over. This is very probably due to 
the fact that in the measuring conditions around April the approximation of the spectral 
lines by gaussians is not good enough and it would need a better treatment.

Therefore, in what follows, we will concentrate in the observations made with the null-measurement
method which, by definition, are done in the most favourable conditions,  
do not depend on the calibration of the measurements and, anyway, they coincide very well
with the noon-measurement method (see Fig.\ref{Figcali}).

\begin{figure}
\center
\includegraphics[width=\columnwidth]{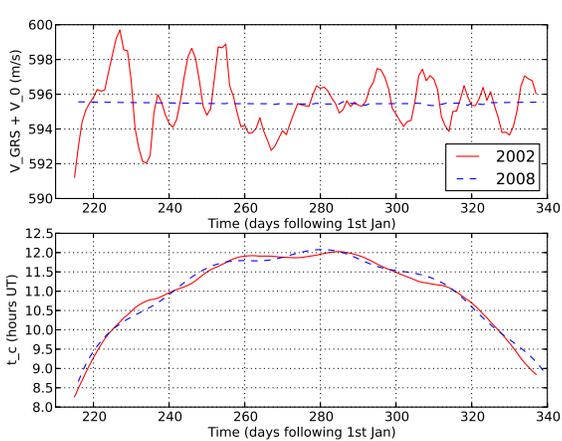} 
\caption{\label{Figvgrsimul} Results from the numerical model (see text) for years 2002 (near maximum 
 solar activity) and 2008 (near minimum solar activity). Top graph shows the GRS velocity obtained
 using the same procedure as the one used with the observations, while 
on the bottom graph the crossing time is showed. 
Both parameters are obtained with the same procedure
applied to the observed data (see section \ref{sec:nullGRS}).
}
\end{figure}

When the numerical simulation is applied to predict the GRS measurements at the croosing time (null-measurement
method), the results found
 can be seen in Fig. \ref{Figvgrsimul} for the same years than before (2002 and 2008). 
The bottom graph shows the crossing time,
obtained using the same procedure as the one used for the observations, where a small 
 variation, with a period of nearly a month,
can be clearly seen, due to the residual effect of the Moon on the Earth's orbit. 
The top graph shows
the result of the GRS velocity variations for both years. Notice that, at solar 
activity minimum, the variation is negligible being  well below 1 m s$^{-1}$ while at maximum activity
a signal of up to $\approx$ 8 m s$^{-1}$ peak to peak variation is seen.

\begin{figure}
\center
\includegraphics[width=\columnwidth]{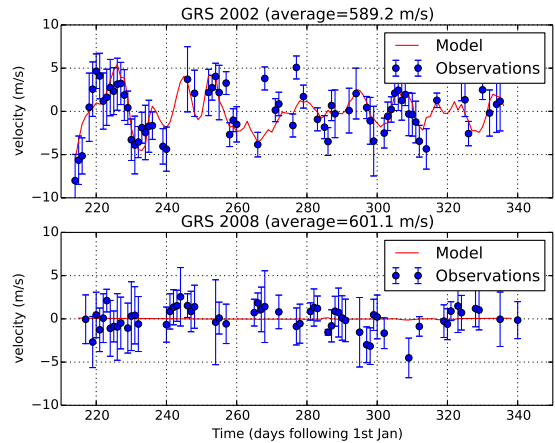}
\caption{\label{Figvgrsanual_sim} Comparison of the GRS velocity results obtained from 
observations (blue dots) and the results of the numerical simulation 
(red line) in a year close to solar activity maximum (2002, top) and for another close 
to minimum solar activity(2008, bottom).
}
\end{figure}

\begin{figure*}
\center
\includegraphics[height=22.6 cm]{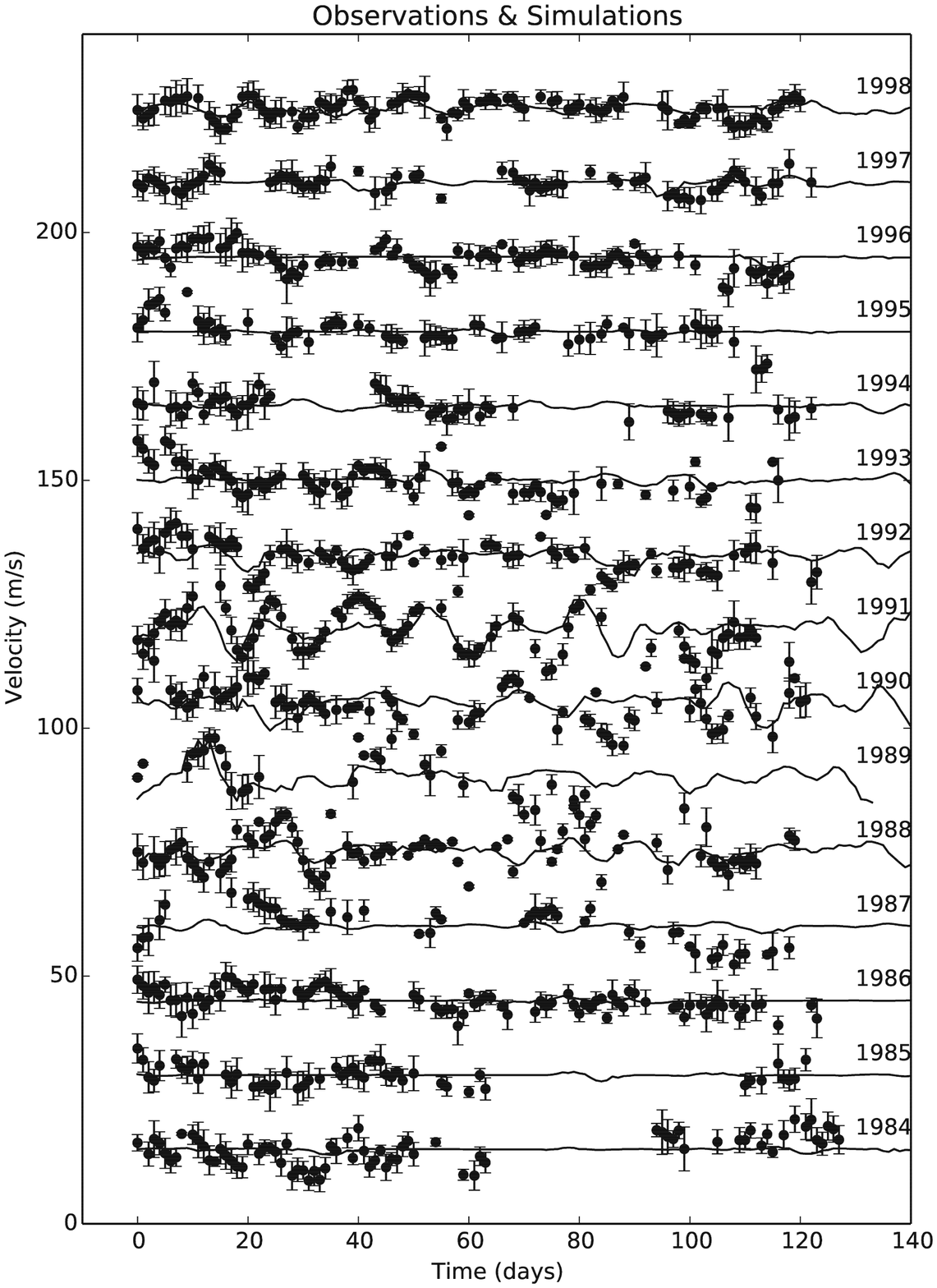} 
\caption{\label{Figrefer1} Comparison of the observations and the results obtained from the numerical model 
described in the text.The graph shows the annual observed GRS velocity (with its average substracted) and 
the results obtained from the numerical model. For clarity, the vertical scale is built by adding 15 m/s to 
previous years' results. The first day plotted is the first day of that year with null-method measurements.
}
\end{figure*}

\begin{figure*}
\center
\includegraphics[height=22.6 cm]{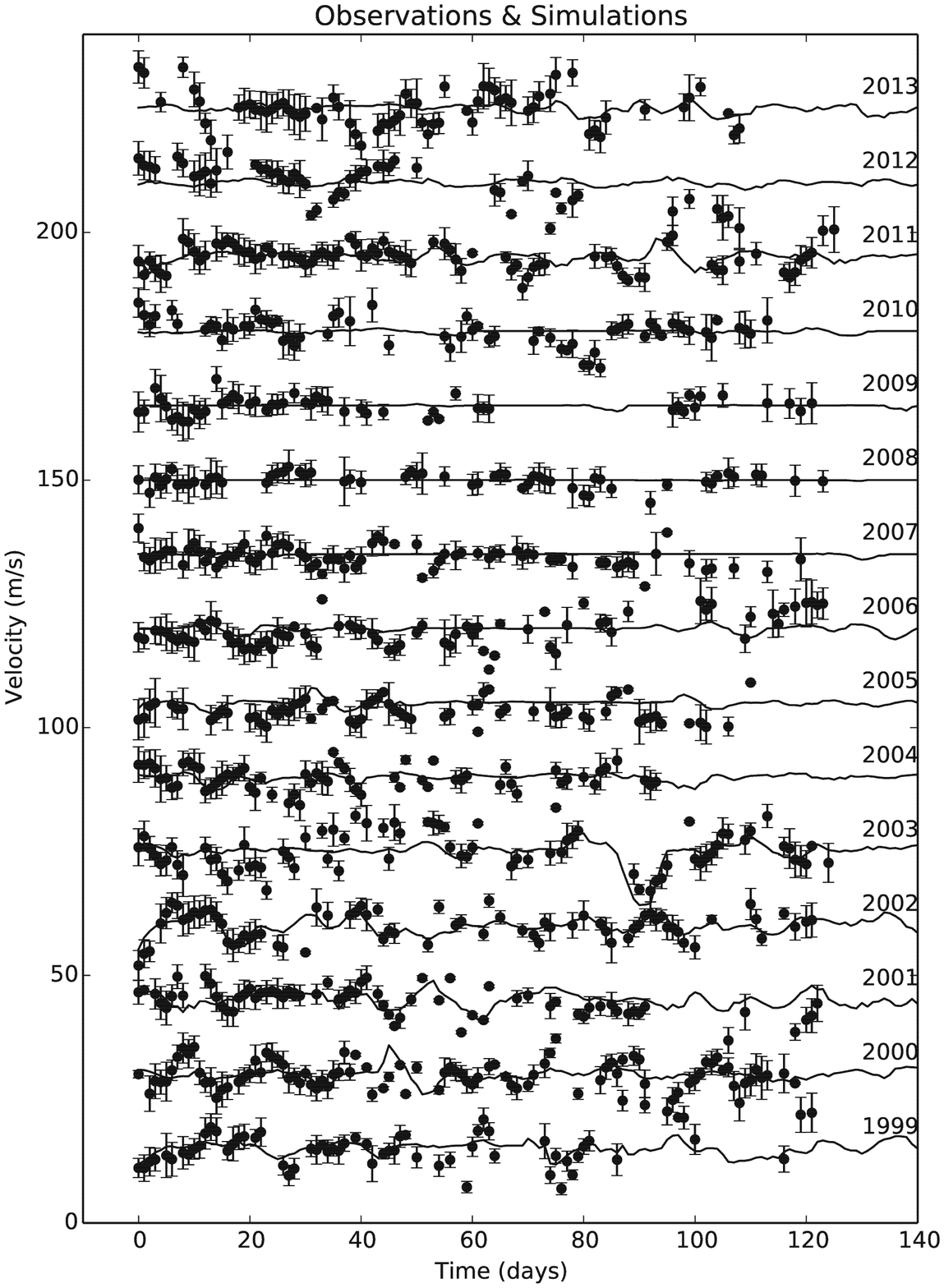} 
\caption{\label{Figrefer2} Comparison of the observations and the results obtained from the numerical model 
described in the text.The graph shows the annual observed GRS velocity (with its average substracted) and 
the results obtained from the numerical model. For clarity, the vertical scale is built by adding 15 m/s to 
previous years' results.The first day plotted is the first day of that year with null-method measurements.
}
\end{figure*}

In this case, the comparison of the numerically simulated results with observations can be 
seen for the same years in Figure~\ref{Figvgrsanual_sim}. The match is very good in 2002 
given the observational errors, with the exception of 
certain details. For instance, the amplitude of the simulated signal is somewhat 
smaller than the observed one even though they both follow the same trends.
In some other years, the simulated signal seems sometimes  to lead the 
observed one (as already pointed out in \citep{Herretal1984}). These details would 
suggest that more realism in the simulation would be interesting
and necessary, with the inclusion 
of the contribution from faculae, for instance, but this would be difficult owing to the lack of 
sufficient observational data for these features. Moreover, the inclusion of a better 
approximation of the line profiles (solar and lab) would probably help to obtain a closer match
of the amplitudes. 
In Fig.\ref{Figrefer1} and Fig.\ref{Figrefer2} the results obtained for the 
years 1984 to 2013 are shown. Observations prior to 1984 were only limited summer campaigns 
and were performed with different heliostat system and readout electronics (see section \ref{obsproc}).
 
\subsection{Instrumental systematic errors}
\label{sec:corinstrum}

A thorough analysis of the possible instrumental errors in the apparatus 
is to be found in \cite{BIR1978} and in \cite{Teo1979}. During the 
observing days many tests have been done that confirm the expected 
results of these analyses. Such tests were always performed by changing 
one parameter at a time and leaving at least a complete day of observation to 
be able to measure the potential effect of the change on the observed data. 
Moreover, from 1985 onwards, some housekeeping data, such as temperatures
at different points of the spectrophotometer (allways kept below 1K fluctuations) 
and guiding errors, have also
been recorded. These are very small, for an integration time of 40 seconds the error is
1.8$\pm$0.6 arcsec and it is below 0.1 arcsec for any given day. The pointing sensitivity 
has been measured to be less than 0.8 m s$^{-1}$/arcmin.

\begin{figure}
\center
\includegraphics[width=\columnwidth]{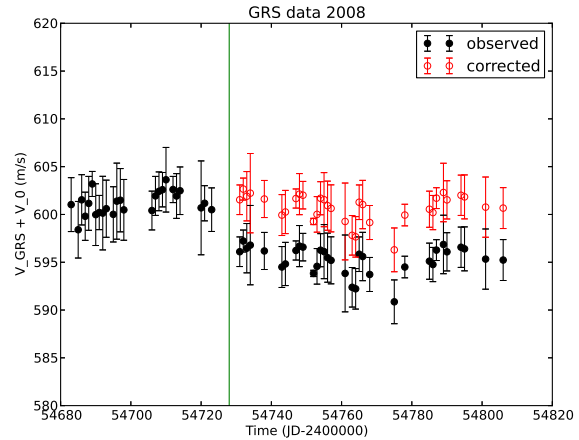} 
\caption{\label{Figmounts} Instrumental effect on the observed GRS velocity due to 
the summer to winter secondary mirror mount changeover (a green 
vertical line shows the precise day) and the corresponding re-alignement. Notice that 
if a positive shift is added to the 
black dot points following the mount changeover's day  to obatin the red open circle points, 
the observed velocity now shows a nice continuity with the observations obtained 
before.
}
\end{figure}

To summarize the results here, it can be said that the systematic 
errors (mechanical, thermal, optical and electronic) are kept well 
below the 1 m s$^{-1}$ level in all parts of the instrument. 
Possible electronic assymetry between
both channels (L and R) and long term instrumental stability were 
checked in dedicated long runs using appropriate calibration set-ups; 
all tests have shown uncertainties of 
$\approx \pm$0.1 m s$^{-1}$. Moreover, the magnetic field of the permanent magnet
showed no significant change over the years. However, the most 
sensitive source of error is misalignment of the electro-optical 
light modulator (EOLM) with respect to the entrance beam. 
This can happen near the equinoxes, when the secondary 
mirror mounting has to be changed. This is a major change and the alignment is checked, 
a realignement 
usually being required; however, the same relative position between the 
entrance beam and the EOLM axis is sometimes not achieved, resulting in a slight change
in the offset velocity. The resulting 
effect on the daily offset velocity measurement is illustrated in 
Fig.\  \ref{Figmounts} with an example; in the graph, the day of the mount 
changeover is shown by a green vertical line, and a step in the 
measured velocity can be seen. By calculating the mean 
value of a few points before and after the mount changeover we can correct for such systematic effects.

Another systematic effect might come from the 
existence of the isotope $^{41}K$ with a proportion of 
7.4\% respect to  $^{39}K$ which produces an isotopic 
shift of up to $-$98 m s$^{-1}$ as the optical thickness of the vapour 
changes from thin to thick \citep{Jackson1938}. These effects 
will be measurable if the optical thickness of the gas is different 
 in the cell from that on the Sun's atmosphere, or if it changes with time. Moreover, while there 
is no reason for a significant change in optical thickness on the Sun, 
in the cell it can change only if the temperature of the oven heating the cell 
changes. This parameter has been measured (every minute) and recorded from
1985 onwards; it is found to vary very smoothly during the day 
typically by $\pm$ 1 degree (which represents 0.7\%) owing 
to room temperature variations. The same cell was used 
from 1976 to 2009; in the years 1977, 1985 and 2000  tests were performed
by carefully changing the setting of the oven voltage heating circuit by an amount 
 as high as 10\% (in power), with very small effects of less than 0.5 m s$^{-1}$/\% being found in the 
 results.

\section{Discussion of the results}

\begin{figure}
\center
\includegraphics[width=\columnwidth]{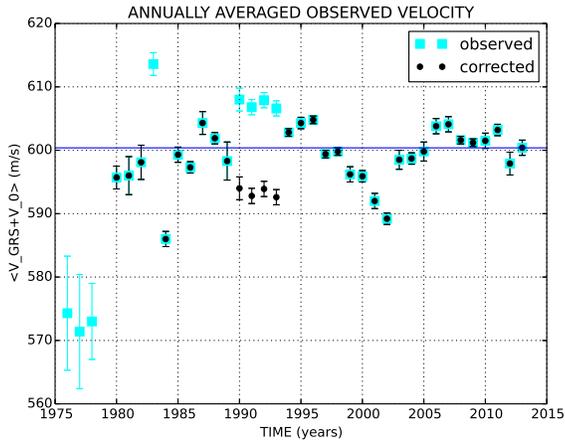}
\caption{\label{Figvgrsobser}  Yearly averages of the observed 
velocities (cyan squares) following the subtraction of the solar 
activity effect on the measurements. 
In black circles, the same data once 
 some values (1990--93) for instrumental systematic errors are corrected
and others (1976--1978,1983) due to different heliostat 
set-up (see text) are removed.
}
\end{figure}

Once the numerically calculated effect of the passage of 
the sunspots is corrected, yearly averages of the GRS velocity 
are obtained; these are plotted in Figure~\ref{Figvgrsobser}. As can be 
seen, there are some points that diverge considerably from the mean value such 
as those obtained during the first three years; data 
from 1976 to 1978 result some 25 to 30 m s$^{-1}$ lower than the rest. This 
is interpreted as an instrumental effect brought about by the fact that, in these years, the heliostat 
system that feeds the spectrophotometer with the incident beam was 
different (the primary was displaced from the north--south line defined 
by the secondary and had to be changed at noon)
from that used afterwards and included mirrors of much lower quality.
Moreover, in 1983 we had very few useful results (as much as six) as the 
electronic channel analyser in the data collection system did not  work 
properly and had to be replaced. These points 
have therefore been disregarded. Finally, in the years 1990 to 1993, the systematic effect in 
the secondary mount changeover was kept 14 m s$^{-1}$ too high. Once these considerations were 
taken into account and the data points were corrected, 
a second definitive graph (see black dots in Fig.~\ref{Figvgrsobser}) 
was produced.

\begin{figure}
\center
\includegraphics[width=\columnwidth]{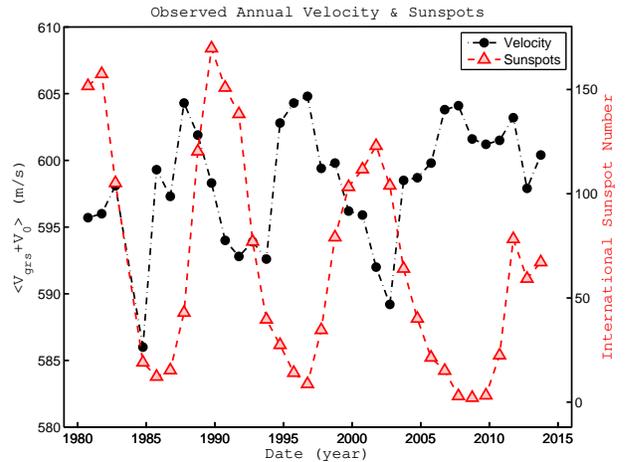} 
\caption{\label{Figtotal}  Final yearly averages of the GRS observed 
velocity (dots) and averages of the international sunspot number 
(triangles) calculated over the same time interval than the observations. 
}
\end{figure}

These changes have been considered in the data and the final average found is  600.38 $\pm$ 0.78 m s$^{-1}$,
which corresponds to 94.75 $\pm$ 0.1 \% of the full effect predicted by the principle of equivalence. 
Note that the statistical error is the lowest ever measured on the Sun. However, 
it also shows a variation of  $ \approx \pm$ 5 m s$^{-1}$ with a period of 10--12 
years, which is very close to the solar activity cycle's mean periodicity. In order 
to understand this effect we have plotted, in Fig.\ \ref{Figtotal}, 
the results found, together with an average of the international sunspot 
number over the same days where we had GRS data, taken from \citep{SIDC2014}
(http://sidc.oma.be/sunspot-data/). A clear anti-correlation 
can be seen between the two graphs, suggesting an explanation in terms of 
magnetic activity influencing the symmetry of the solar line profile.

Such a 5\% difference from the predicted GRS value can be explained as follows. 
The only instrumental error that
can account for such difference is that in the  optical depth of the potassium 
atoms in the Sun rather than in the cell; if this were so, the differential isotopic shift could explain
the difference. However, in our opinion, this is not the most plausible explanation. 
The solar line is a strong resonance line that is optically thick in the Sun. In the laboratory the vapour cell,
in the MarkI apparatus, the optical depth depends on the number of atoms present in its head, 
which is a consequence of the temperature at which the cell is being heated. The voltage in the oven
is supplied by a very stable power supply that heats the cell up to 360 K; the oven voltage is measured
with a thermistor, which shows that the temperature is kept stable close to 0.1 K during observations.
The temperature is set so that a plateau in the scattered intensity is achieved; in these conditions
the vapour is optically thick in the cell's head. However, \cite{Snider1974} in a 
research note reports the measurement of 16.4 $\pm$ 1 m\AA ,  which corresponds to 638 $\pm$ 39 m s$^{-1}$
using a potassium-based atomic beam resonant scattering spectrometer. Given that the error is much higher than ours,
the difference in the value found here can be explained because in our case we are measuring integral sunlight
rather than using a small aperture in the centre of the Sun's image, therefore using
a differently averaged solar potassium line profile and also a differently averaged solar limb shift effect.

The limb shift effect of this solar line, and the GRS itself, have also been measured by \cite {LoPresto1994}
at the Mac Math-Pierce solar telescope also using a small aperture on a large solar image.
They also measured 99  $\pm$ 7 \% of the full predicted GRS value 
and a small linear limb shift. Another measurement of this effect has also
been made by \cite {Andersenetal1985}, who found a non-linear limb shift 
increasing rapidly when approaching the solar limb. The 
difference between both seems to be in the amount of scattered light 
from the centre of the Sun in the measurements at positions
approaching the limb; in both works, the position of the solar line 
is determined by appropriately fitting its minimum. A third measurement came from
our own group \citep {Angueraetal1987}, which found an intermediate result. 
Moreover, in our numerical simulation (see subsection \ref{sec:corsolares}) 
we find that the effect averaged over the whole Sun is $\approx$ 15 m s$^{-1}$.

The resonant scattering experiments measure the solar line in its wings rather than
by finding the minimum. Therefore, if the line is not symmetric the measurement 
can be either blue- or redshifted with respect to the minimum. 
Measurements of the asymmetry of the KI7699 \AA\  line
made by \cite {LoPresto1994} and \cite {Bonetal1988} show that,
at between 30 and 60\% of the residual intensity, it is blueshifted at the centre of
the Sun, while close to the limb it is centred or even 
redshifted with respect
to the minimum. Using integral sunlight, \cite {Rocaetal1983} find that the wings 
are blueshifted  by $\approx$ 50 $\pm$ 30 m s$^{-1}$ with respect to the minimum.
Very similar results are obtained if we analyze the data on the KI7699\AA\  line from
the integral sunlight atlas taken at Kitt Peak \citep{solaratlas1984}.

Taking such considerations into account, we believe this latter explanation,
asymmetry of the solar line (-50 m s$^{-1}$) plus residual limb shift effect (15 m s$^{-1}$), explains
the average GRS measured value.

\section{Conclusions}

The solar gravitational redshift measured 
in the KI7699 \AA\  line for the years 1976 to 2013 has been 
obtained as a by-product of the measurements of 
solar oscillations taken on integral sunlight at Observatorio de El Teide,
from 1976 to 2013, using the resonant scattering spectrometer MarkI.

The average result over this period 
results in 600.4 $\pm$ 0.8 m s$^{-1}$, with a clear variation, which is in 
anti-correlation with the phase of the solar activity cycle, of roughly $\pm$ 5 m s$^{-1}$. 
Although the statistical error bar is the lowest ever measured, correction for 
this effect should bring it even lower. 
These measurements are amongst the most precise ever made of the 
solar gravitational redshift and the most numerous.

The value found for the GRS is 33 m s$^{-1}$ lower than the theoretically 
value expected (5\% of the full effect). We believe that, besides considerations already taken into account,
the anti-correlation found between the variations around the mean value 
and the solar activity cycle also suggest that the discrepancy can be 
attributed to the overall assymetry of the Sun-integrated KI7699\AA\  
spectral line changing with magnetic activity. 

An obvious consequence that can also be derived is that, when
 the radial velocities of other stars are studied, the currently reported 
findings warn against interpretations such as possible planets or
 other components orbiting around them, without first checking for stellar 
 magnetic activity effects.
 
 To improve this result it would be very convenient to avoid the use of solar 
magnetic sensitive lines. Moreover, 
the use of integral sunlight prevents observational problems, such as diffuse light,
specially when measuring near the limb. However, the determination of the overall 
effect on integral sunlight of the convective limb-shift, in order to 
properly correct the measurements, is a very difficult task. Therefore, the use of spectral lines
with small limb-shift would also be advisable.
%
%
\small  
%
\section*{Acknowledgments}   

The data used in this work, their temporal coverage spanning almost 40 years (an 
astronomer's lifetime), their continuity and quality, have involved 
the contribution of many individuals who developed and maintained the apparatus, the site where
it stands and performed the observations, and also the support of several 
governmental funding institutions from Spain and the UK. It all started forty years ago, through a collaboration between
the HiRES group at the Physics Department of the University of Birmingham (UB, UK) and the solar section of the 
Instituto Universitario de Astrof\'{i}sica at the Universidad de La Laguna (ULL, Spain). 
We remain deeply indebted to all the individuals who made it possible.

However, this paper is dedicated to the memory of our friends and colleagues, the late 
Bill Brookes, George Isaak and Bob van der Raay (and their families) from 
the University of Birminghan (UB), actual 
designers, builders and early sustainers of the MarkI spectrophotometer apparatus. 
Also to the memory of the late Joan Casanovas (who began the collaboration), 
Montse Anguera and Irene Gonz\'alez from the Instituto de Astrof\'{i}sica de Canarias (IAC). 
One more detail is that the topic of this paper was first suggested and encouraged
by Prof. G. R. Isaak.

We are also thankful to our technical colleagues who took care of updates and the main repairs of the 
instrument:  Clive McLeod, Brek Miller and Steven Hale (and the late Joe Litherland) at UB 
and,  Ezequiel Ballesteros and Andrew Jones at the IAC.  Special mention 
of the maintenance team of the 
Observatorio del Teide (IAC) lad by Ignacio del Rosario and the administrative 
and supportive personnel lead by Miquel Serra.
We also thank the personnel of the Servicios Inform\'aticos Comunes (SIC) of the IAC. 
 The MarkI instrument manpower requirements at 
the beginning and the end of each daily 
observing run implies the need for observers at the site.
Over the past 40 years, many people, nearly 100, have performed the observations: 
our colleagues and staff scientists of the IAC, summer students at the IAC, 
undergraduate students from the Faculty of Physics at the ULL 
and, recently, the team of observers at the Observatorio del Teide, coordinated by Alex Oscoz. 
We acknowledge their diligence and dedication. Many thanks to LLuis Tom\'as, Ricard Casas, 
Santiago L\'opez and particularly to 
Antonio Pimienta, who has been for more than twenty years (still he is) the 'curator'
of the instrument and of its environment at the lab.

Staff scientists of the IAC's Helioseismology group have played a 
crucial role in the Mark-I data scientific exploitation, as most of them began their 
scientific career with it: Juan C. P\'erez, Gary Bramford, Clara R\'egulo, Antonio Jim\'enez, 
Fernando P\'erez, Stuart  Jefferies, Luis S\'anchez,
Jes\'us Patr\'on, Isabel Mart\'{i}n, Antonio Eff-Darwich, Rafa Garc\'{i}a, Chano Jim\'enez and other colleagues
that helped with the observations. Other institutions at the site, like KIS (Kieppenheuer Institut f\"ur Sonnenphysik)
always helped aluminizing the mirrors. We are also indebted to Manuel V\'azquez, the former leader of the incipient 
Solar Physics group at the IAC, for his continuous support.
The Mark-I spectrophotometer became in the mid-nineties a node of the BiSON 
project (Birmingham Solar Oscillation Network), which consisted of another 
four nodes under the responsibility of Prof.\ Yvonne Elsworth (UB). 

Institutional support have provided the services and funding necessary to ensure the 
continuity of MarkI observations: we thank the UB 
and the IAC for providing so many services (administration, electronic and mechanical 
workshops amongst others) over the entire operational period of the  Mark-I. The Spanish Government, 
under different funding schemes that evolved during these years (currently the Spanish 
National Plan of Research and Development under grant AYA2012-17803), and British 
institutions (SERC at the beginning, then PPARC and, at present, the STFC, 
Science and Technology Facilities Council) have supported the MarkI project and 
later the whole BiSON network. Currently, the whole MarkI database is accessible at the SVO site 
(Spanish Virtual Observatory  http://svo2.cab.inta-csic.es/vocats/marki); 
this initiative was developed in the framework of the FP7-SPACE-2011-1, project n.312844 (SPACEINN).

\end{document}